\documentclass[a4paper]{EslabStyle}
\usepackage{graphics}
\usepackage{psfig}

\begin{document}

\title{Transitional YSOs: Candidates from Flat-Spectrum IRAS Sources}

\author{A.W. Volp\inst{1} \and E.A. Magnier\inst{2,3} \and
        M.E. van den Ancker\inst{1,4} \and
        L.B.F.M. Waters\inst{1,5}}
\institute{
Astronomical Institute ``Anton Pannekoek'', Kruislaan 403, NL-1098 SJ 
Amsterdam, The Netherlands \and
Astronomy Dept. 351580, University of Washington, Seattle, WA 98195, USA \and
Canada-France Hawaii Telescope, P.O. Box 1597, Kamuela, HI 96743, USA \and
Harvard-Smithsonian Center for Astrophysics, 60 Garden Street, MS 42, 
Cambridge, MA 02138, USA \and
Instituut voor Sterrenkunde, Katholieke Universiteit Leuven, Celestijnenlaan 
200B, B-3001 Heverlee, Belgium}

\maketitle 

\begin{abstract}
We are searching for Young Stellar Objects (YSOs) near the boundary between 
protostars and pre-main sequence objects, what we have termed transitional 
YSOs. We have identified a sample of 125 objects as candidate transitional 
YSOs on the basis of IRAS colors and optical appearance on DSS images. 
We find that the majority of our objects are associated with star-forming 
regions, confirming our expectation that the bulk of these are YSOs.

We present optical, near-IR and high-resolution IRAS images of 92 objects 
accessible from the northern and 62 from the southern hemisphere. The 
objects have been classified on the basis of their morphology and spectral 
index. Of the 125 objects, 28 have a variety of characteristics very similar 
to other transitional YSOs, while another 22 show some of these 
characteristics, suggesting that these transitional YSOs are not as rare 
as predicted by theory.
\keywords{Circumstellar Matter -- Stars: Formation -- 
Stars: Pre-Main Sequence -- ISM: Jets and outflows}
\end{abstract}

\section{Introduction}
The transition between a Class I and a Class II source (Lada \& Wilking 1984)
is one of the less well-known phases in the life of a Young Stellar Object 
(YSO). This period is also one of the most interesting in the evolution of 
a young star as outflow phenomena, which may determine the final mass 
of the star and process the material in the surrounding molecular 
cloud, are particularly active at this stage. However, only few 
objects are known at or near this boundary between protostar and 
pre-main sequence star, limiting our possibilities to gain more 
insight in the physical mechanisms behind this rapid transition. 
Recently a nebulous object, Holoea (IRAS 05327+3404; Hawaiian for 
flowing gas), was discovered which shows some characteristics of a Class I 
source (flat spectrum, outflow), but also has some Class II characteristics 
(optically visible central star). This object has increased its optical 
brightness over the last 50 years, suggesting that it is in the process 
of becoming exposed and making the transition between a protostar and 
a pre-main sequence star (Magnier et al. 1996, 1999). Guided by the 
observed properties of Holoea, we have therefore performed a systematic 
search for additional candidates for the group of transitional YSOs, 
the results of which are presented in these proceedings.
\begin{table*}[bht]
\caption{Transitional YSO Candidate IRAS Sources}
\label{tab:table}
\begin{center}
\scriptsize
\begin{tabular}{ll@{\hspace{2.0mm}}l@{\hspace{2.0mm}}ll}
\multicolumn{3}{l}{Category 1 Identifications:}\\
\noalign{\smallskip}
\hline\noalign{\smallskip}
IRAS ID      &     RA (J2000) DEC       & notes \\
\noalign{\smallskip}
\hline\noalign{\smallskip}
00294$+$6510 & 00 32 18.5  $+$65 27 19  & one very red star, bright neighbour, refl. neb. \\
00353$+$6249 & 00 38 17.1  $+$63 06 01  & one very red star, refl. neb. \\
03260$+$3111 & 03 29 10.4  $+$31 21 58  & {\it in NGC 1333}, very red star, lots of refl. neb. \\
03383$+$4343 & 03 41 44.8  $+$43 52 54  & red star, some neb. \\
03507$+$3801 & 03 54 05.5  $+$38 10 39  & {\it by refl. neb. PP 11}, red star\\
04115$+$5027 & 04 15 22.2  $+$50 34 37  & very red star, several m. red stars \\
04553$-$6921 & 04 55 05.3  $-$69 16 55  & {\it in LMC}, some neb., busy field \\
05327$+$3404 & 05 36 05.4  $+$34 06 11  & {\bf Holoea!}, {\it in M36, NGC 1960},  very red star + refl. neb. \\
05373$+$2349 & 05 40 24.5  $+$23 50 53  & {\bf CPM 19 YSO},{\it in KOY98 81} 1 very red star \\
06047$-$1117 & 06 07 08.3  $-$11 17 51  & a very red star + neb. (emis?)\\
06244$+$0336 & 06 27 02.5  $+$03 34 21  & very red star \\
06567$-$0350 & 06 59 14.5  $-$03 54 51  & {\bf BFS 56},{\it in FT96 217.4-0.1} very red star, neb.\\
06568$-$1154 & 06 59 13.0  $-$11 58 56  & {\bf CMa West} \\
06584$-$0852 & 07 00 51.6  $-$08 56 28  & {\bf CPM 33 YSO},{\it in FT96 221.9-2.0}, red stars, neb. \\
08211$-$4158 & 08 22 52.3  $-$42 07 56  & {\bf HH obj},{\it in vdB 15}, refl. neb.\\
13224$-$5928 & 13 25 40.6  $-$59 43 42  & {\bf YSO },{\it in DCld 307.3+02.9},1 very red star, neb.\\
14563$-$6301 & 15 00 24.9  $-$63 13 34  & {\it in vdB 65}, 1 red star, neb. \\
15064$-$6429 & 15 10 40.9  $-$64 40 28  & {\bf NGC 5844, PK 317-5.1 PN} Plan. neb.\\
15365$-$5435 & 15 40 21.0  $-$54 45 00  & red star with cometary neb. \\
17340$-$3757 & 17 37 29.6  $-$37 59 22  & very red star, ext. emis. neb. \\
18018$-$2426 & 18 04 53.8  $-$24 26 40  & {\bf RAFGL 2059}, {\it in M8E region, by S25}, very red star, ext. emis. neb. \\
20024$+$3330 & 20 04 22.5  $+$33 38 58  & {\bf G070.7+01.2 (many IDs)}, Some controversy... \\
20193$+$3448 & 20 21 18.7  $+$34 57 48  & very red star + neb. \\
20236$+$4058 & 20 25 27.8  $+$41 08 19  & {\it in LBN 253}, very red star + neb. \\
20337$+$4036 & 20 35 32.7  $+$40 46 33  & {\it in LBN 271}, very red star \\
20582$+$7724 & 20 57 13.1  $+$77 35 46  & {\it in L 1228}, dark neb., several red obj, neb. \\
21569$+$5842 & 21 58 36.4  $+$58 57 08  & {\it in L 1143}, very red star, neb. \\
23395$+$6358 & 23 41 56.0  $+$64 15 09  & a single very red star \\
\noalign{\smallskip}
\hline \\
\multicolumn{3}{l}{Category 2 Identifications:}\\
\noalign{\smallskip}
\hline\noalign{\smallskip}
IRAS ID      &     RA (J2000) DEC       & notes \\
\noalign{\smallskip}
\hline\noalign{\smallskip}
02259$+$7246 & 02 30 43.8  $+$72 59 39  & {\it in L1340}, faint red star, refl. neb.  \\
04020$+$5017 & 04 05 47.0  $+$50 25 07  & several (2-3) m. red stars, no obvious neb. \\
04038$+$5437 & 04 07 50.1  $+$54 45 33  & several (2-3) m. red stars, no obvious neb. \\
04278$+$2435 & 04 30 52.7  $+$24 41 49  & {\bf ZZ Tau YSO}, by mol. cl. OMK96 30, 1 m. red star \\
05223$+$1908 & 05 25 16.3  $+$19 10 45  & one red star, some neb. \\
05343$+$3605 & 05 37 41.8  $+$36 07 20  & {\it by S233, S231}, several m. red stars, 1 very red + neb. \\
06041$+$3012 & 06 07 23.8  $+$30 11 44  & {\bf MWC 790 HAeBe? }1 very red star, is cluster? \\
06535$+$0037 & 06 56 06.0  $+$00 33 51  & {\bf CPM 31 YSO, ZOAG 212.96+01.29}, m. red star, neb. \\
07166$-$1816 & 07 18 50.8  $-$18 22 11  & some neb. \\
07183$-$2741 & 07 20 21.1  $-$27 47 02  & {\bf Bran 19 },one red star, some neb.  \\
07221$-$2544 & 07 24 13.6  $-$25 50 03  & {\it in Bran 23 },one red star, some neb. (emis?)  \\
07254$-$2259 & 07 27 35.0  $-$23 05 25  & some neb.\\
07466$-$2631 & 07 48 43.4  $-$26 39 29  & some neb., spike from HD 63599\\
08404$-$4033 & 08 42 17.1  $-$40 44 10  & {\bf ESO H$\alpha$ 162},{\it in BRAN 174}, refl. neb.\\
08500$-$4254 & 08 51 49.2  $-$43 05 30  & {\it in star forming region?}, red star, some faint neb.\\
10381$-$5704 & 10 40 09.0  $-$57 20 03  & one red star, some neb.\\
19025$+$0739 & 19 04 60.0  $+$07 44 24  & several red stars, no neb. \\
19050$+$0524 & 19 07 32.7  $+$05 29 41  & {\it by S74}, several m. red stars, dark neb.? \\
19365$+$2557 & 19 38 34.6  $+$26 04 47  & one red star, no neb, globule? \\
20078$+$3528 & 20 09 44.7  $+$35 37 05  & {\it in LBN 182}, diff. neb, several m. red stars\\
20172$+$3554 & 20 19 10.7  $+$36 03 54  & one very red star, some m. red stars, no neb. \\
22206$+$6333 & 22 22 18.0  $+$63 48 51  & {\it in L 1204}, some red stars \\
\noalign{\smallskip}
\hline \\
\multicolumn{3}{l}{Category 4 Identifications:}\\
\noalign{\smallskip}
\hline\noalign{\smallskip}
IRAS ID      &     RA (J2000) DEC       & notes \\
\noalign{\smallskip}
\hline\noalign{\smallskip}
05017$+$2639 & 05 04 50.6  $+$26 43 18  & {\bf HD 32509}, bright star, some faint neb. HAeBe?\\
05293$+$1701 & 05 32 14.2  $+$17 03 25  & {\bf HD 36408}, pair of bright stars, highly sat. \\
06303$+$1021 & 06 33 04.4  $+$10 19 20  & {\bf NGC 2247 nebula}, sat in g, i, J, K. neb? \\
15532$-$4210 & 15 56 42.5  $-$42 19 25  & {\bf HD 142527 HAeBe} \\
18585$-$3701 & 19 01 55.3  $-$36 57 11  & {\bf R CrA HAeBe}, {\it in NGC 6729 dif. neb.}, very lum. refl. neb. \\ 
19111$+$0212 & 19 13 41.7  $+$02 17 39  & {\bf PK 37-3.3 symb}, bright star \\
19340$+$2228 & 19 36 09.6  $+$22 35 14  & {\bf HD 184961}, bright star \\
\noalign{\smallskip}
\hline
\end{tabular}
\end{center}
\end{table*}

\section{Selection Criteria}
Our initial selection criterion for transitional YSO candidates is based 
on their infrared spectrum. We selected all point sources with infrared colors 
similar to Holoea ($-1.3 < [12]-[25] = +2.5 \log (f_{12}/f_{25}) < -0.40$, 
$-2.0 < [25]-[60] = +2.5 \log (f_{25}/f_{60}) < -1.0$) and reliable 
data in the 12, 25 and 60 $\mu$m bands (Category 3 detections) from the 
IRAS Point Source Catalog. This resulted in a list of 327 IRAS sources. 
To narrow down this list, we examined small (4\arcmin $\times$ 4\arcmin) 
fields extracted from the Digital Sky Survey (DSS) in centered on each 
source to check for traces of nebulosity, as is also seen near Holoea. Since 
the nuclei of Seyfert galaxies may have IRAS colors similar to our selected 
range, a number of sources which were clearly associated with spiral galaxies 
were also rejected. This resulted in 125 candidate sources. The vast majority 
of these were found to be located in the vicinity of CO clouds and generally 
near other signs of active star formation. This lends credence to our 
suggestion that the bulk of these sources are young stellar objects.

\section{Observations}
For the southern sources in our list of 125 transitional YSO 
candidates, CCD imaging in the Bessel $V$, Bessel $R$ and Gunn $i$ bands 
was carried out on the Dutch 90cm telescope at ESO, La Silla. For the 
sources accessible from the north, CCD images of each source in 
$g_*$, $r_*$ and $i_*$ (Krisciunas et al. 1998) and near-infrared 
$J$, $H$ and $K^{\prime}$ filters were taken at the University of 
Washington Apache Point Observatory 3.5m telescope, New Mexico. 
Data were reduced in a standard fashion, after which they were 
positionally and flux-calibrated and aperture photometry was 
performed for the dominant optical source in the IRAS error ellipse. 
In addition to this, we generated high-resolution IRAS 60 $\mu$m 
maximum entropy processed images (HIRAS; Bontekoe et al. 1994) of 
the environment of each candidate.

\section{Results}
\begin{figure*}[ht]
\centerline{\psfig{figure=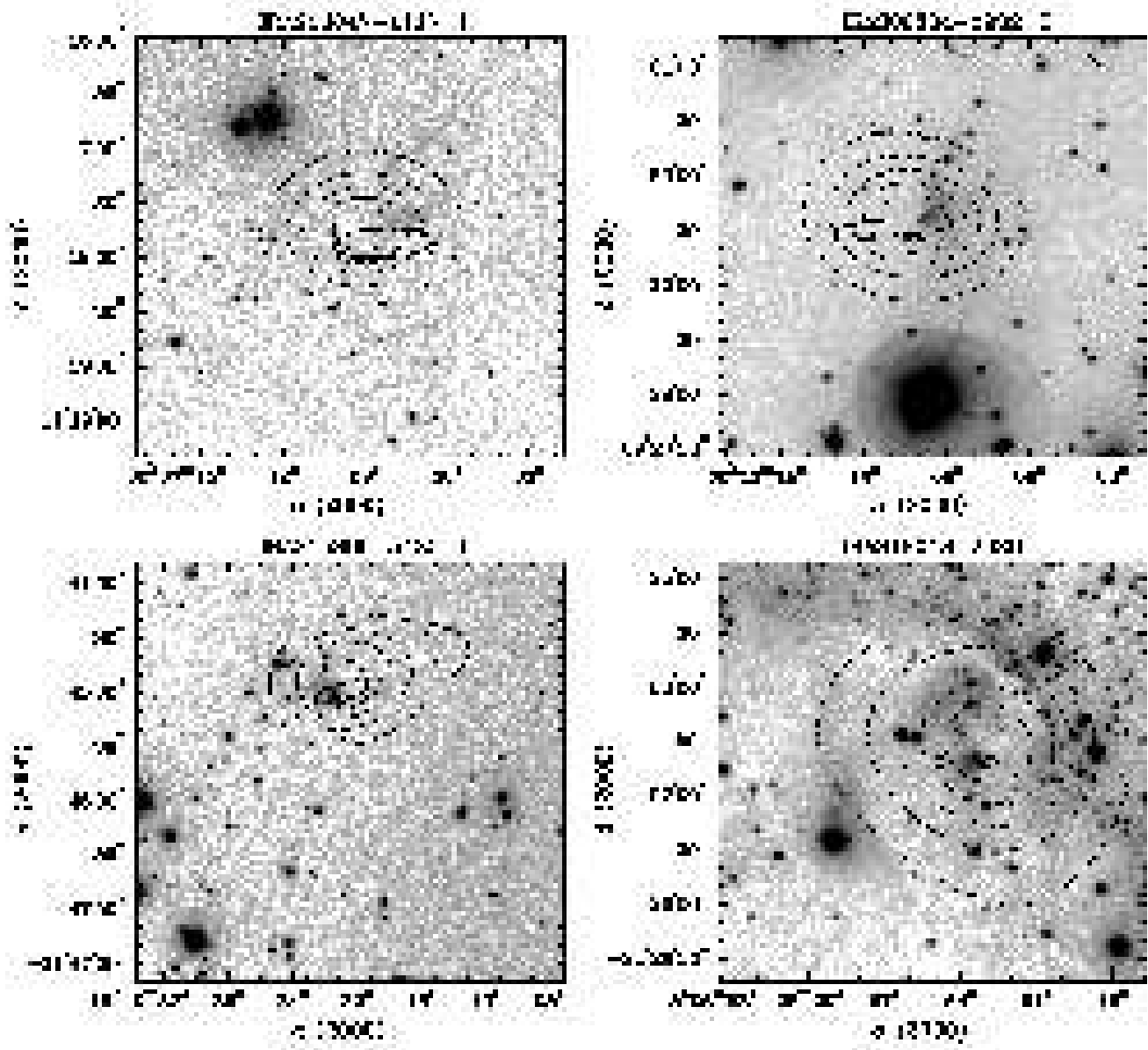,width=17.2cm,angle=0}}
\caption{Examples of $i$ band images of southern sources obtained 
         with the Dutch telescope at La Silla. From left to right, top 
         to bottom: IRAS 06047$-$1117, IRAS 06384+0932, IRAS 15365$-$5435 
         and IRAS 18018$-$2426. Also shown are the 
         error ellipse from the IRAS Point Source Catalogue (solid 
         line) and contours from the HIRAS 60~$\mu$m image (dashed 
         lines). \label{fig1}}
\end{figure*}
We have classified the objects in seven categories based on their 
morphology in the CCD images and the spectral index of the
dominant optical source in the IRAS error ellipse:
\begin{enumerate} 
\item A likely transitional YSO:  A single moderately-bright, very red
  stellar object with extensive associated reflection nebulosity. (28~objects)
\item A possible transitional YSO: a moderately red stellar object
  with weak nebulosity or a significantly red object with no
  nebulosity. (22~objects)
\item A YSO group: Several very red objects, usually with extended
  nebulosity.  No single object stands out. (21~objects)
\item Bright (Herbig Ae/Be like) star. (7~objects)
\item A cluster of stars: usually a red cluster with no single very
  red star. (18~objects)
\item A galaxy. (11~objects)
\item Nothing: no object stands out, and no object can be associated
  with any of the other classes. (18~objects)
\end{enumerate}
The sources with Category 1 and 2 classifications (likely transitional 
YSOs) and Category 4 classifications (Herbig Ae/Be star candidates) 
are listed in Table~1.
\begin{figure}[ht]
\centerline{\psfig{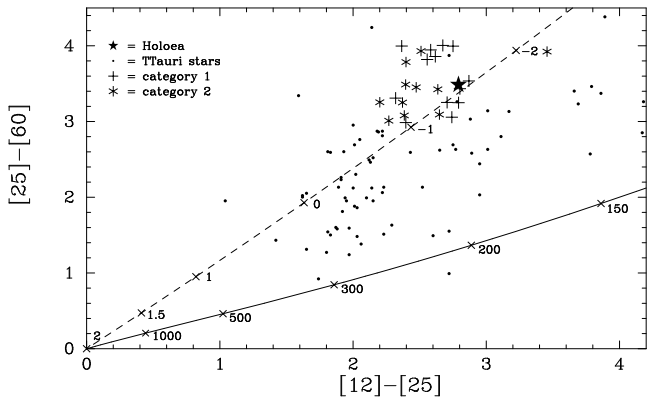}}
\caption{IRAS [12]-[25] vs. [25]-[60] color-color diagram of our category 
         1 (crosses) and 2 (asterisks) sources. For comparison we also 
         show the location of Holoea (star), the IRAS colors of T Tauri 
         stars in Taurus-Auriga (Kenyon \& Hartmann 1995; dots), the location 
         of blackbodies of different temperatures (solid line) and the 
         location of different power-law spectral indices (dashed line). 
         \label{fig2}}
\end{figure}
\begin{figure}[ht]
\centerline{\psfig{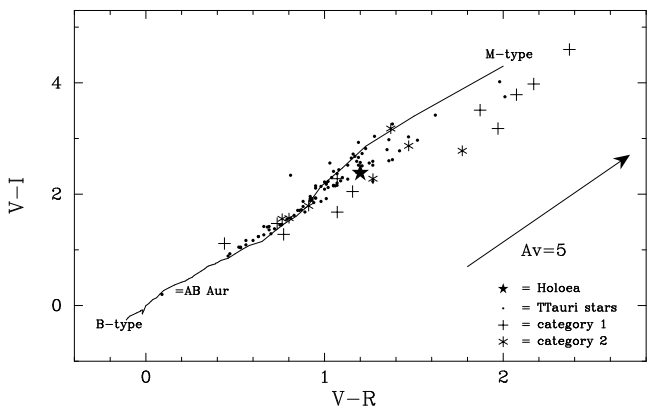}}
\caption{$V-R$ vs $V-I$ color-color diagram of our category 
         1 (crosses) and 2 (asterisks) sources. For comparison we also 
         show the location of Holoea (star), the colors of T Tauri 
         stars in Taurus-Auriga (Kenyon \& Hartmann 1995; dots), and the 
         colors of main sequence stars (solid line). The large arrow 
         shows the direction of normal interstellar reddening.
         \label{fig3}}
\end{figure}

In Figures 2 and 3 we show infrared and optical color-color diagrams 
of the sources in our sample with Category 1 and 2 identifications. For 
comparison we also show the location of T Tauri stars in Taurus-Aurigae, 
taken from Kenyon \& Hartmann (1995) in these diagrams. On average our 
Category 1 and 2 sources have redder IRAS colors than T Tauri stars. They 
have an infrared energy distribution that is wider than a blackbody, with 
a typical peak temperature of 200--300~K. The slope of their energy 
distribution appears compatible with infall rather than a steady-state 
disk. 

The optical colors of the Category 1 and 2 sources in our sample are 
on average also redder than those of T Tauri stars. Because the 
slope of the reddening direction in the $V-R$ versus $V-I$ diagram 
is nearly parallel to that of decreasing effective temperature, we 
cannot determine with the current data whether this is caused by a 
difference in stellar temperature or due to additional extinction 
in our sources. However, the mere fact that these sources are on 
average very red suggests that our identification of the optical 
source with the IRAS source is correct and these objects do 
indeed have an energy distribution that is similar to Holoea.

\section{Conclusions}
Our selection criteria to identify transitional YSO candidates have 
been rather successful. Of the 125 objects, 28 have a variety of 
characteristics very similar to other transitional YSOs, while another 
22 show some of these characteristics. The fact that our Category 1 and 
Category 2 YSO candidates show on average redder optical and infrared 
colors than T Tauri stars agrees with our hypothesis that these objects 
are in the process of making the transition between Lada Class I and II.
If confirmed, this would suggest that these transitional YSOs are not as 
rare as predicted by theory.
In addition to this, we have found seven objects to be good candidates 
for members of the Herbig Ae/Be stellar group, of which three are newly 
identified as such. A follow-up study of our Category 1, 2 and 4 YSO 
candidates using newly obtained optical spectroscopy and submillimeter 
spectral-line data is under way and will allow us to make a more 
detailed assessment of the nature of these sources.

\begin{acknowledgements}
EAM acknowledges support by NWO/Astron under contract number 
782-376-011.  Support for EAM was also provided by NASA through
grant number GO-06459.01-95A from the Space Telescope Science
Institute.  LBFMW acknowledges financial support through a NWO 
{\it Pionier} grant.  MvdA acknowledges financial support from 
NWO grant 614.41.003.
\end{acknowledgements}

\end{document}